\documentclass[amstex,onecolumn,showpacs,floats,floatfix,aps,pra]{revtex4}
\usepackage{amsfonts}
\usepackage{amssymb}
\usepackage{amsmath}
\usepackage{graphicx}
\usepackage{bm}
\usepackage{color}
\usepackage{epsfig}
\usepackage{ifthen}

\input{tcilatex}

\begin{document}

\author{A. A. Rangelov}
\affiliation{Department of Physics, Sofia University, James Bourchier 5 blvd., 1164
Sofia, Bulgaria}
\title{Factorizing numbers with classical interference: several
implementations in optics}
\date{\today}

\begin{abstract}
Truncated Fourier, Gauss, Kummer and exponential sums can be used to
factorize numbers: for a factor these sums equal unity in absolute value,
whereas they nearly vanish for any other number. We show how this
factorization algorithm can emerge from superpositions of classical light
waves and we present a number of simple implementations in optics.
\end{abstract}

\maketitle

\section{Introduction\label{Introduction}}

Factorization of numbers into their prime factors is a hard non-polynomial
problem for classical computers. It was Shor \cite{Shor algorithm} who
proposed a quantum algorithm which can solve the problem of factorization of
numbers on a quantum computer with a tremendous speedup as compared to a
classical computer. A practical demonstration of Shor's algorithm has been
carried out by factorizing the integer 15 \cite{factorizing 15}, using
nuclear magnetic resonance. However, quantum computers capable of
implementing Shor's algorithm for larger numbers have not been developed yet.

Several approaches to factorize numbers based on interference of multiple
quantum paths have been proposed \cite%
{Dowling,Summhammer,Merkel2006,Mahesh,Mehring,Gilowski2008,Bigourd2008,Sadgrove2008,Stefanak}%
. Those schemes do not use quantum entanglement and do not capitalize on
quantum parallelism. As a consequence, these schemes scale exponentially
with the number of digits of the factorized number. This is in contrast to
Shor's algorithm which requires only a polynomial number of operations.
Nevertheless, if the interference is implemented in a suitable way in a
system, which does the factorization, then one can benefit because nature
plays the role of a computer.

As was pointed out by Jones \cite{Jones} the proposed techniques for
factorization based on Gauss sums \cite{Mehring,Mahesh,Gilowski2008}
unfortunately do not provide useful methods to factorize numbers, because a
precalculation of the factors is needed for the experiment. In spite of
that, Gauss sums would be useful if it is possible to avoid explicit
precalculation stages of the algorithms.

Physical systems that can implement the Gauss sums must be described by
complex numbers. In the present paper we investigate how truncated Fourier
sum and its generalizations, like truncated Gauss, Kummer and exponential
sums, could emerge from superposition of several oscillations. Those sums
can be used successfully to factorize numbers. Due to the wide use of
interferences and beats in optics, we shall keep our consideration close to
optics and our examples are in wave optics too.

However, the proposed implementation can be extended to virtually any
physical system where superposition among several different oscillations
appear, from the mechanical pendulum with several degrees of freedom,
through the atomic and solid state systems and their analogs in quantum
mechanics.

\section{Truncated Fourier, Gauss, Kummer and exponential sums}

In order to find the factors of a given number $N$ we use the following
truncated sum:
\begin{equation}
\mathcal{A}_{N}^{(M)}(l)=\frac{1}{M}\sum_{m=1}^{M}\exp \left( -2\pi im^{k}%
\frac{N}{l}\right) ,  \label{truncated sum}
\end{equation}%
where $k$ is an integer and $M$ is the number of terms in the sum. The
argument $l$ scans through all integers between $1$ and $\sqrt{N}$ for
possible factors. The capability of the sum of Eq. (\ref{truncated sum}) to
factor numbers originate from the fact that for an integer factor $q$ of $N$
with $N=ql$, all phases in $\mathcal{A}_{N}^{(M)}(l)$ are integer multiples
of $2\pi $. Consequently, the terms add up constructively and yield $%
\mathcal{A}_{N}^{(M)}(l)=1$ . When $l$ is not a factor, the phases oscillate
rapidly with $m$, and $\mathcal{A}_{N}^{(M)}(l)$ takes on small values. In
this interference pattern, larger truncation parameter $M$ leads to better
convergency. In principle, already the first several terms of the sum are
sufficient to discriminate factors from non-factors. Depending on the
coefficient $k$ in Eq.(\ref{truncated sum}) we distinguish several important
cases:
\begin{subequations}
\label{Fourier, Gauss and Kummer}
\begin{gather}
\text{Fourier sum for }k=1\text{\cite{Stefanak},} \\
\text{Gauss sum for }k=2\text{ \ \cite%
{Merkel2006,Mahesh,Mehring,Gilowski2008,Sadgrove2008,Stefanak,Merkel2007,Stefanak2007}%
,} \\
\text{Kummer sum for }k=3\text{ \cite{Stefanak},} \\
\text{exponential sum for }k=m\text{ \cite{Stefanak}.}
\end{gather}

The use of quadratic phases to factor numbers (Gauss sum) has the advantage
of fewer terms needed in the sum to distinguish factors from non-factors
compared to the linear phase (Fourier sum), which is because of high
quasi-randomness for the quadratic phase \cite{Stefanak}. In the very same
way the Kummer sum, and sums with nonlinear phases of higher order, has an
advantage compared to the Gauss sum \cite{Stefanak}.

Now we consider a system with $M$ different oscillation modes, with
frequencies $\omega _{m}$, phases $\varphi _{m}$ and amplitudes $E_{0m}$
\end{subequations}
\begin{equation}
E_{m}\left( t\right) =E_{0m}\exp \left( i\omega _{m}t+i\varphi _{m}\right) .
\label{individual oscillation}
\end{equation}%
Using the superposition principle we can write the resulting oscillation as
the sum of all oscillations:
\begin{equation}
E\left( t\right) =\sum_{m=1}^{M}E_{m}\left( t\right)
=\sum_{m=1}^{M}E_{0m}\exp \left( i\omega _{m}t+i\varphi _{m}\right) .
\label{result oscilation}
\end{equation}%
In the sum of Eq. (\ref{result oscilation}) we can vary the parameters $%
E_{0m}$, $\omega _{m}$, $\varphi _{m}$ and the time $t$. In the next several
sections we will show how truncated Fourier, Gauss, Kummer and exponential
sums could emerge when we fix three of the parameters for all oscillations,
while changing the fourth parameter.

\section{Factorization using differences in time delay (interferometry)}

\subsection{Mach-Zehnder interferometer}

First we consider the case when the parameters $E_{0m}$, $\omega _{m}$, $%
\varphi _{m}$ are equal for all oscillations in Eq. (\ref{individual
oscillation})
\begin{eqnarray}
E_{0m} &=&E_{0}, \\
\omega _{m} &=&\omega , \\
\varphi _{m} &=&0,
\end{eqnarray}%
thus the only parameter that is left not fixed in Eq. (\ref{individual
oscillation}) is the time $t$. This can be easily realized in optics by
interferometry, where the individual oscillations describe the electric
field for the different arms of the interferometer as shown in Fig.\ref{four
arms}.

\begin{figure}[t]
\begin{center}
\includegraphics[width=180mm]{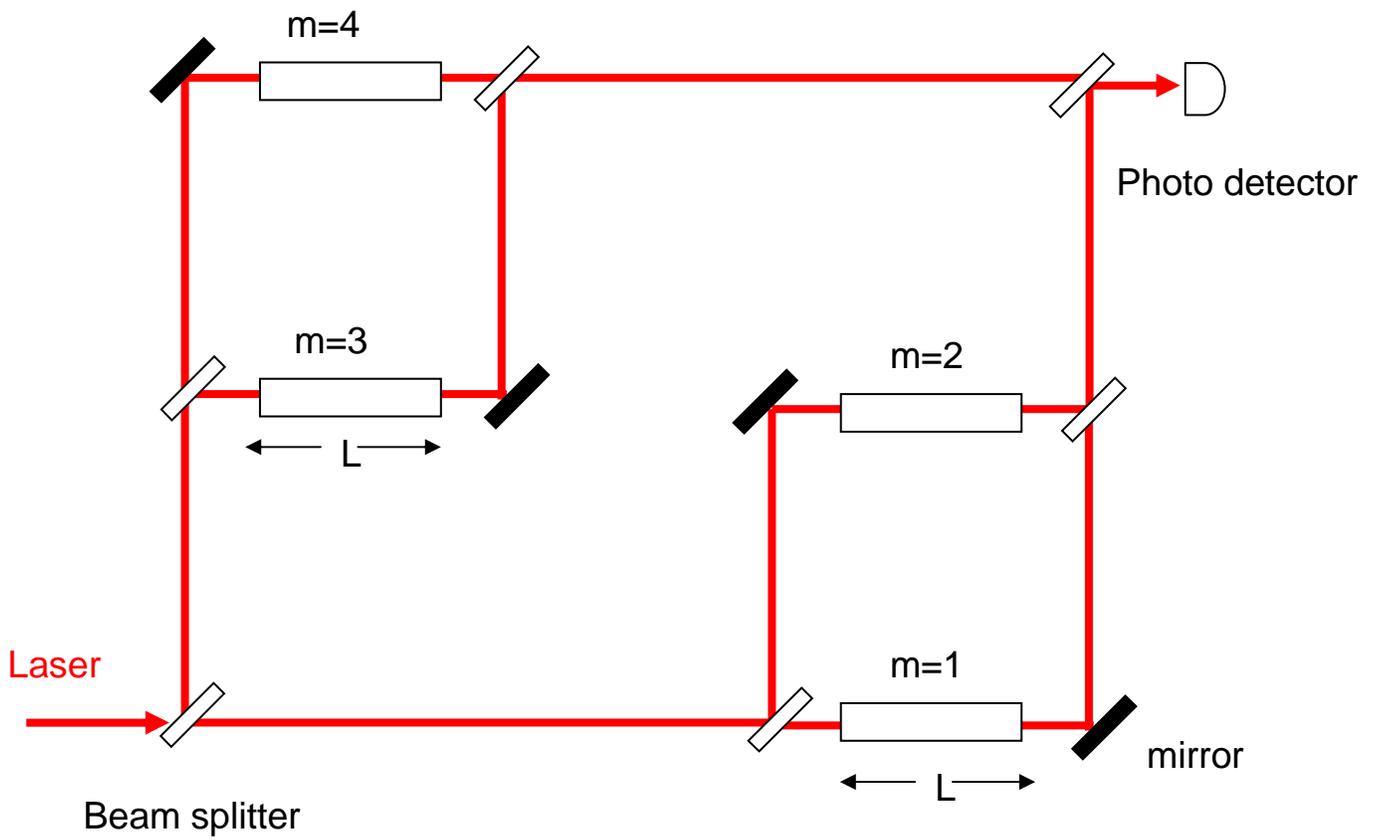}
\end{center}
\caption{Four arms Mach-Zehnder interferometer that can be used to factorize
numbers by using Fourier, Gauss, Kummer or exponential sums. The four arms
correspond to four terms in the sum. Repeating the procedure for doubling
the arm in principle one can increase the terms in the sum as mach as needed
to distinguish factors from nonfactors.}
\label{four arms}
\end{figure}

From Fig. \ref{four arms} and Eq. (\ref{result oscilation}) we see that we
have the following sums of electric fields in the detector%
\begin{equation}
E=E_{0}\sum_{m=1}^{M}\exp \left( i\phi _{m}\right) ,
\end{equation}%
where $\phi _{m}$ is the phase accumulated in the $m$ arm of the
interferometer due to the difference in travel time through each arm.

Suppose that each arm of the interferometer is with length $L,$ the wave
length of the light that we use in vacuum is $\lambda $ and the
corresponding frequency is $\omega $, let the index of refraction in each
arm of the interferometer is different and is denoted as $n_{m}$. Then the
phase $\phi _{m}$ for the beam that travels through the $m$-th arm of the
interferometer is given as%
\begin{equation}
\phi _{m}=t_{m}\omega =\frac{L}{c_{m}}\omega ,
\end{equation}%
here $t_{m}$ is the time that light travel in the $m$-th arm of the
interferometer to pass length $L$ and $c_{m}$ is the speed of light in that
arm. The refraction index in arm $m$ of the interferometer is
\begin{equation}
n_{m}=\frac{c}{c_{m}},
\end{equation}%
thus
\begin{equation}
\phi _{m}=\omega \frac{L}{c}n_{m}=\frac{2\pi }{\lambda }n_{m}L,
\end{equation}%
then the electric filed in the detector is
\begin{equation}
E=E_{0}\sum_{m=1}^{M}\exp \left( 2\pi i\frac{L}{\lambda }n_{m}\right) .
\end{equation}%
Now if the index of reflection in the $m$ arm of the interferometer is
\begin{equation}
n_{m}=a+bm^{k},
\end{equation}%
then
\begin{equation}
E=E_{0}\exp \left( 2\pi i\frac{aL}{\lambda }\right) \sum_{m=1}^{M}\exp
\left( 2\pi im^{k}\frac{bL}{\lambda }\right) .
\end{equation}%
The detector registers the intensity,%
\begin{equation}
I\sim \left\vert E\right\vert ^{2}=\left\vert \sum_{m=1}^{M}\exp \left( 2\pi
im^{k}\frac{bL}{\lambda }\right) \right\vert ^{2}.
\end{equation}%
Various $k$ gives us a different type of truncated sum (see Eq. (\ref%
{Fourier, Gauss and Kummer})). The number that we want to factorize is $bL,$
the trial factors are $\lambda $. Each time when the trial factor $\lambda $
is a factor of $bL$ we will observe a maximum signal in the detector. The
number of the terms in the sum can be controlled by doubling the elements in
the interferometer Fig. \ref{four arms}. The numbers that could be
factorized in this way are of order $L/\lambda \sim \frac{1m}{1000nm}=10^{6}$%
.

\subsection{Pulse train}

Now we consider a train of pulses, where the delay of the $m$ pulse compared
to first pulse is given as%
\begin{equation}
t_{m}=m^{k}\tau ,
\end{equation}%
here $m$ takes the values $m=1,2,3...M$, while $\tau $ can be set as a unit
of time.

We consider the case when all pulses have equal amplitudes $E_{0m}=E_{0}$
and equal frequencies $\omega _{m}=\omega $. Then the electric field for the
$m$ pulse is given by Eq. (\ref{individual oscillation}) and reads
\begin{equation}
E_{m}=E_{0}\exp \left( i\omega t_{m}+i\varphi _{m}\right) =E_{0}\exp \left(
i\omega m^{k}\tau +i\varphi _{m}\right)
\end{equation}%
Let us make a different path way for every pulse in such a way that all
pulses hit the same detector at the same time, this is equivalent to make $%
\varphi _{m}=0$ at the place where all pulses collides. Then the intensity
that the detector registers is a result from the superposition among all
electric fields, e.g the sum from Eq. (\ref{result oscilation}):%
\begin{equation}
I\sim \left\vert E\right\vert ^{2}=\left\vert \sum_{m=1}^{M}E_{m}\right\vert
^{2}=\left\vert E_{0}\sum_{m=1}^{M}\exp \left( 2i\pi m^{k}\nu \tau \right)
\right\vert ^{2},  \label{Intensity}
\end{equation}%
where $\nu =\omega /\left( 2\pi \right) $. If one chooses the frequency $\nu
$ as the number that we want to factorize ($N$) and $1/\tau $ as a trail
factor ($l$), then Eq.(\ref{Intensity}) reduces to the sum from Eq. (\ref%
{truncated sum}).

\section{Factorization using differences in frequencies (beats)}

If we now consider a system that exhibits several oscillations with the same
amplitude $E_{0}$ and the same initial phases ($\varphi _{m}=0$), but with
different frequencies $\omega _{m}$, then the individual oscillations (\ref%
{individual oscillation}) are described by
\begin{eqnarray}
E_{m}\left( t\right) &=&E_{0}\exp \left( i\omega _{m}t\right) , \\
\omega _{m} &=&m^{k}\omega _{0},  \label{the angular frequency}
\end{eqnarray}%
the resulting oscillation (\ref{result oscilation}) is
\begin{equation}
E\left( t\right) =\sum_{m=1}^{N}E_{0}\exp \left( -2\pi im^{k}\nu
_{0}t\right) ,
\end{equation}%
where $\nu _{0}=\omega _{0}/\left( 2\pi \right) $. We will observe beats
when $t\nu _{0}$ is a integer which could be used to find the factors of \
the number $\nu _{0}$. One physical realization of the above idea could be a
light with several high-harmonic generated frequencies \cite%
{Brabec2000,Gavrila1992}, chosen in the way that they present for example
the odd terms in the Fourier sum:
\begin{equation}
\omega _{0},3\omega _{0},5\omega _{0},7\omega _{0}....
\end{equation}%
then in the detector the time of the detection play the role of the test
factors and whenever there is a beat we observe a maximum of the signal,
thus this time is a real factor.


\section{Factorization using Faraday effect}

The last parameter that we can vary in Eq. (\ref{individual oscillation}) is
the amplitude of the individual oscillation. For example if we work with
laser light we can use the different polarization orientations of the
electric field. The electric field is a vector in the polarization plane,
which can be described by complex electrical field.

\begin{figure}[t]
\begin{center}
\includegraphics[width=170mm]{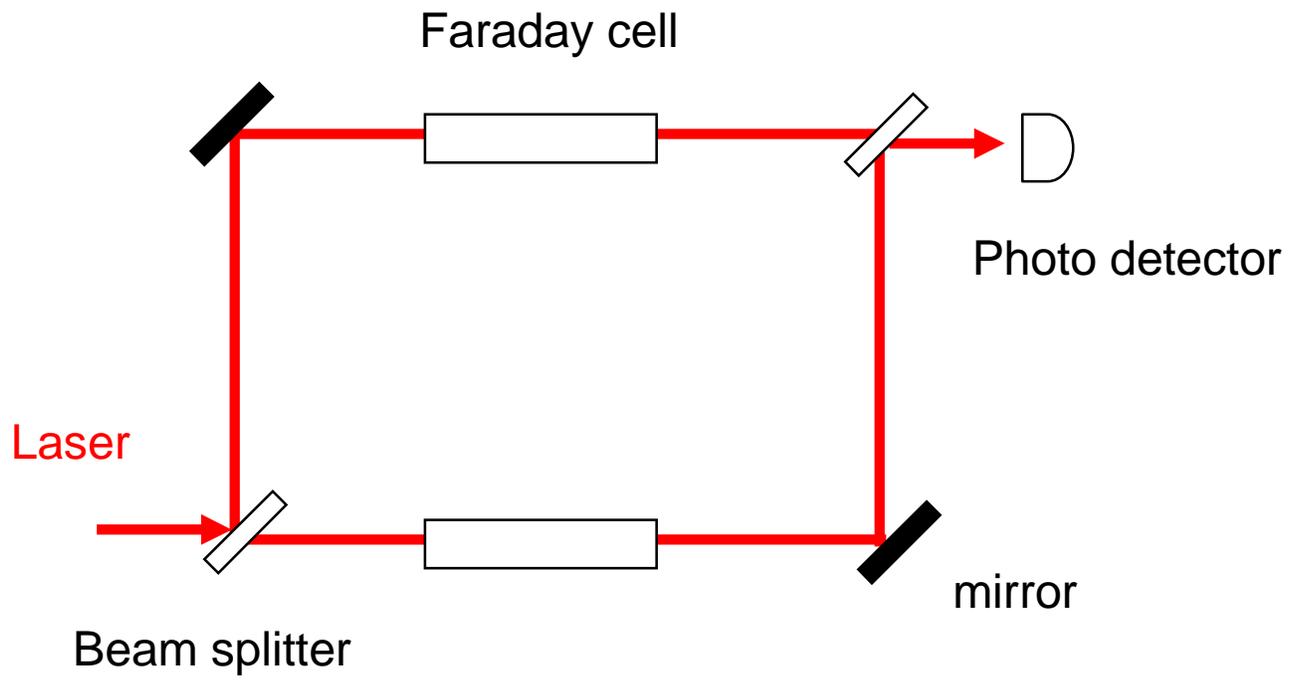}
\end{center}
\caption{Two different path ways of the light that pass through Faraday
cells and are used to factorize numbers.}
\label{Faraday cells}
\end{figure}

Let us consider the case when we have a linearly polarized light pulse,
which is split in several parts and each part passes different pathways
through Faraday cells as shown in Fig. \ref{Faraday cells}.

Applying different Faraday rotation angles $\varphi _{m}$ on each pathway
and collecting all of the light at the same place (at the detector) the
resulting electric field is the superposition:%
\begin{equation}
E=\frac{E_{0}}{M}\sum_{m=1}^{M}\exp \left( i\varphi _{m}\right) ,
\end{equation}%
where $E_{0}$ is the electrical field amplitude of the initial beam. The
relation between the angle of polarization rotation due to the Faraday
effect $\varphi _{m}$ and the magnetic field $B_{m}$ in a diamagnetic
material \cite{Lifshitz2000} is

\begin{equation}
\varphi _{m}=2\pi bLB_{m},
\end{equation}%
where $L$ is the length of each pathway and $2\pi b$ is the Verdet constant
for the material \cite{Lifshitz2000}. For the amplitude of the resulted
electric field we have:%
\begin{equation}
E=\frac{E_{0}}{M}\sum_{m=1}^{M}\exp \left( 2\pi ibLB_{m}\right) .
\end{equation}%
If we now have a magnetic field $B_{m}$ for the $m$-th Faraday cell, which
is given as:
\begin{equation}
B_{m}=B_{0}m^{k},
\end{equation}%
then the intensity in the detector is
\begin{equation}
I\sim \left\vert E\right\vert ^{2}=\left\vert \frac{E_{0}}{M}%
\sum_{m=1}^{M}\exp \left( 2\pi im^{k}bLB_{0}\right) \right\vert ^{2}.
\end{equation}%
Here the number that we want to factorize is $bL,$ the trial factors are $%
1/B_{0}$.

\section{Conclusions}

We have shown how the factorization algorithm based on truncated Fourier,
Gauss, Kummer or exponential sums emerges naturally from superpositions of
classical light waves. We have proposed a number of simple implementations
in optics. These implementations can be extended to virtually any physical
system where superpositions of several different oscillations appear.

The factorization algorithms discussed in this paper are classical
algorithms and thus their complexity scales exponentially with the number of
digits. If an extension of this algorithm exists in entangled quantum
systems, then a quantum computing parallelism would be involved with an
exponential speedup of factorization. The present solutions therefore could
be the first step to an alternative quantum factorization algorithm to the
famous Shor algorithm.

\acknowledgments
This work has been supported by the EU ToK project CAMEL, the EU RTN project
EMALI, the EU\ ITN project FASTQUAST, and the Bulgarian National Science
Fund Grants No. WU-2501/06 and No. WU-2517/07. The author is grateful to N.
Vitanov for stimulating discussions and critical reading of the manuscript.
During the preparation of this paper, the author became aware of a related
work by Tamma et al. \cite{Tamma2008}.

\bigskip
\bigskip
\bigskip

\newpage

\bigskip
\bigskip
\bigskip

\end{document}